\documentclass[%
    preprint,
    superscriptaddress,
    amsmath,
    amssymb,
    aps,
    pre,
    table, 
    xcdraw
]{revtex4-1}

\usepackage{amssymb}
\usepackage{setspace}
\usepackage[utf8]{inputenc}
\usepackage{array}
\usepackage{natbib}
\usepackage{amsmath}
\usepackage{float}
\usepackage{multirow}
\usepackage{hyperref}
\usepackage{rotating}
\usepackage{adjustbox}
\usepackage{xcolor}

\setcitestyle{authoryear,open={(},close={)}}

\begin{document}

\title{Predicting Citation Impact of Research Papers Using GPT and Other Text Embeddings}

\author{Adilson Vital Jr.}
\affiliation{Institute of Mathematics and Computer Sciences,
University of Sao Paulo, PO Box 369,
13560-970, S\~ao Carlos, SP, Brazil 
}

\author{Filipi N. Silva}
\affiliation{Indiana University Network Science Institute, Bloomington, IN, USA
}

\author{Osvaldo N. Oliveira Jr.}
\affiliation{Sao Carlos Institute of Physics, PO Box 369,
University of Sao Paulo, PO Box 369,
13560-970, S\~ao Carlos, SP, Brazil 
}

\author{Diego R. Amancio}
\affiliation{Institute of Mathematics and Computer Sciences,
University of S\~ao Paulo, PO Box 369,
13560-970, S\~ao Carlos, SP, Brazil 
}

\date{\today}

\begin{abstract}
\textcolor{black}{The impact of research papers, typically measured in terms of citation counts, depends on several factors, including the reputation of the authors, journals, and institutions, in addition to the quality of the scientific work. In this paper, we present an approach that combines natural language processing and machine learning to predict the impact of papers in a specific journal. Our focus is on the text, which should correlate with impact and the topics covered in the research. We employed a dataset of over 40,000 articles from ACS Applied Materials and Interfaces spanning from 2012 to 2022. The data was processed using various text embedding techniques and classified with supervised machine learning algorithms. Papers were categorized into the top 20\% most cited within the journal, using both yearly and cumulative citation counts as metrics. Our analysis reveals that the method employing generative pre-trained transformers (GPT) was the most efficient for embedding, while the random forest algorithm exhibited the best predictive power among the machine learning algorithms. An optimized accuracy of 80\% in predicting whether a paper was among the top 20\% most cited was achieved for the cumulative citation count when abstracts were processed. This accuracy is noteworthy, considering that author, institution, and early citation pattern information were not taken into account. The accuracy increased only slightly when the full texts of the papers were processed. Also significant is the finding that a simpler embedding technique, term frequency-inverse document frequency (TFIDF), yielded performance close to that of GPT. 
Since TFIDF captures the topics of the paper we infer that,  apart from considering author and institution biases, citation counts for the considered journal may be predicted by identifying topics and ``reading'' the abstract of a paper. }
\end{abstract}

\maketitle

\section{Introduction}
\label{chapter:introduction}

The conception of research projects and the selection of a journal for submitting a scientific paper can be influenced by its potential impact, commonly assessed by its anticipated citation count. This impact is determined by various factors, including the journal's reputation, the notoriety of authors~\citep{petersen2014reputation, katchanov2023empirical}, and the quality of the work. The identification of such factors can now be done through data-driven methods~\citep{brito2021associations, brito2023analyzing, fortunato2018science} with the availability of comprehensive datasets, incorporating metadata and citation networks. However, assessing the influence of a paper's content on its impact remains challenging due to the need for an in-depth understanding of its contextual relevance in the field. Techniques for representing and modeling the textual content of documents have been applied to address various complex tasks. These tasks include narrative analysis in social media \citep{georgakopoulou2017small}, topic modeling in academic literature \citep{vayansky2020review}, and automatic summarization and aggregation of news articles \citep{altmami2022automatic, el2021automatic}. In particular, methods such as Generative Pre-trained Transformers (GPT) exhibit remarkable capabilities in enhancing machine translation accuracy \citep{hendy2023good}, generating coherent and contextually relevant text in chatbots \citep{lund2023chatting}, and assisting in advanced predictive text applications \citep{rathje2023gpt}. \textcolor{black}{Since these techniques provide dense and rich representations of complex attributes in textual content, such as topic, writing style, semantics, and even sentiment, one may hypothesize that they can capture attributes related to the impact of a publication}. This would be a crucial addition to comprehending the broader implications of advanced text modeling in academic and research settings, potentially leading to new methodologies for evaluating, understanding, and enhancing the impact of scientific work.

In this paper, we present an approach to predict research impact in terms of the accrued number of citations—based solely on the textual content of papers. Our dataset comprises 40,567 papers from the journal ACS Applied Materials and Interfaces, spanning from 2012 to 2022. To predict citation impact, we employed a variety of text embedding techniques applied to either paper abstracts or full texts. Specifically, we used GPT-based models \citep{DBLP:journals/corr/abs-2005-14165}, the Universal Sentence Encoder (USE) \citep{cer2018universal}, InferSent \citep{DBLP:journals/corr/ConneauKSBB17}, and traditional techniques such as term frequency-inverse document frequency (TFIDF)~\citep{havrlant2017simple} with singular value decomposition (SVD) \citep{stewart1993early}, and Sentence BERT \citep{reimers2019sentence}. These methods generate vector representations of texts, serving as input features for classification or regression models. Our pipeline is aimed at classifying papers based on whether they fall into the top 20\% most cited within the journal. This classification considers two citation metrics: Yearly Citation Count (YCC), which accounts for citations in a given year, and Accumulative Citation Count (ACC), encompassing all citations received post-publication up to a certain year. We used several machine learning algorithms for this task, such as Random Forest \citep{biau2016random}, XGBoost \citep{vital2022comparative,chen2016xgboost}, Logistic Regression \citep{vital2022comparative,menard2002applied}, and kNN \citep{guo2003knn}. In contrast to existing literature, our methodology incorporates (1) various embedding techniques, capturing the semantic richness of abstracts and full text; (2) multiple machine learning models, allowing for a comprehensive evaluation of prediction capabilities; (3) a temporal lens, considering the changing nature of research impact over the years.


\section{Related Works}
\label{chapter:relatedworks}

Several studies have been conducted to predict and analyze scholarly works in science. Examples include prediction of research grant success~\citep{tohalino2021analyzing,tohalino2022predicting,boyack2018toward}, assessment of collaborations among researchers~\citep{guns2014recommending}, and prediction of the success of individual scientists~\citep{acuna2012predicting}. A  prevalent area of investigation revolves around methods to predict citations in academic papers. In~\citep{yan2011citation}, a prediction was made of the number of citations a paper would receive within a specific time frame. This was done mainly with regression models based on features such as topic models, the paper's age measured by the number of years since its publication, diversity in the topics covered, authors' h-index, and other author-related metrics. Additionally, venue centrality was considered. Note that the authors not only utilized features related to the research itself but also those associated with the prestige of researchers and the journals/conferences. The most notable result was an R-squared value of 0.74. The study revealed a relatively precise estimate of long-term citation counts, though it tended to overestimate counts in short-term predictions.

In~\citep{zhao2022utilizing}, the authors introduce a deep learning approach that leverages graph structures and recurrent neural network methods to predict citation counts in academic papers. They utilized the HEP-PH and APS datasets for their analysis~\citep{chen2019information,brito2021associations}. The methodology focuses on using the citation network established shortly after a paper's publication to predict its future impact. This approach is distinct from the current paper in that it solely relies on the citation network, without considering other attributes of the paper, such as its content. The authors observed that their proposed method surpasses other techniques for graph characterization, like node2vec. Deep learning methods appear to effectively preserve the information inherent in citation networks. However, it is also noted that deep learning does not always outperform feature-based methods, more specifically in the APS dataset.

\textcolor{black}{In the study conducted in~\citep{ma2021deep}, the authors used information from the title and abstract to predict the number of citations a paper will receive. They employ Doc2Vec to encode sentences, and features from paragraphs are derived through sentence embeddings using an attention mechanism. The authors reported high accuracy in their predictions and suggested that the prediction could be further enhanced with the use of more recent models.
Unlike our approach, which solely focuses on textual information to predict citations, this paper also considers early citation data. While early citation information proved to be significant, our objective was slightly different. We aimed to predict the impact of a paper as soon as it becomes available, based solely on the paper content, independent of biases that might be associated with the author or journal metadata.
}

\textcolor{black}{
A similar approach that considers textual information to analyze how text can impact the paper impact is described in~\citep{lu2019analyzing}. 
The authors investigated how linguistic complexity, rather than topics, can influence the impact of academic papers. They focused on two disciplines, Biology and Psychology, using data from different sources. The authors probed several variables representing linguistic complexity, such as lexical diversity, lexical sophistication, lexical density, and syntactic complexity, including features like the number of clauses, T-units, and subordination.
Their findings revealed no practically significant relationship between linguistic complexity and impact. Although there were some variations across different disciplines, these differences were not practically meaningful.}

\textcolor{black}{
In a similar study~\citep{porwal2024scientific}, the authors investigated the correlation between scientific impact and linguistic features of papers. They analyzed papers from influential authors in the fields of computer science and electronics, focusing on the titles, abstracts, and conclusions. To measure complexity, they considered various features, including readability, lexical diversity, lexical density, syntactic features, and coherence. Unlike other studies, this paper analyzed the relationship between complexity and impact using correlation methods rather than machine learning. The findings confirmed that linguistic characteristics do play a role in the scientific impact of research papers. However, the influence turned out to be limited. 
}

\section{Methodology}
\label{chapter:methodology}

\subsection{Corpus}

\textcolor{black}{This study analyzed a corpus of 33,803 research papers published in ACS Applied Materials and Interfaces between 2012 and 2020 (inclusive). Citations are available until 2023, meaning a 2020 paper could have at most 3 years of citations. According to OpenAlex~\citep{priem2022openalex}, the papers are divided into four major domains: Physical, Life, Health, and Social Sciences. Physical Sciences has the most papers (31,059), while Social Sciences has the fewest (22). There are also 21 fields (e.g., Materials Science, Engineering, Nursing) and 124 subfields (e.g., Polymers and Plastics, Materials Chemistry, Biomaterials). Table~\ref{tab:papers_per_year} presents the annual publication volume of the journal. Table \ref{tab:paper_distribution} shows the distribution of papers in different fields. Note that the top-cited papers span multiple fields, and this distribution closely mirrors the size distribution of these fields, with a few exceptions. This indicates that their identification is not solely based on topic classification. }
\begin{table}[h]
\centering
\caption{\textcolor{black}{Number of papers published annually in the journal ACS Applied Materials and Interfaces (ACS AMI). `Year' denotes the publication year, and `Papers' indicates the total number of papers published in that particular year.}}
\begin{tabular}{|c|c|}
\hline
\textbf{Year} & \textbf{Papers} \\ \hline
2020 & 5622 \\ \hline
2019 & 5251 \\ \hline
2018 & 5102 \\ \hline
2017 & 4836 \\ \hline
2016 & 4067 \\ \hline
2015 & 3386 \\ \hline
2014 & 2747 \\ \hline
2013 & 1854 \\ \hline
2012 & 938  \\ \hline
\end{tabular}
\label{tab:papers_per_year}
\end{table}

\begin{table}[h]
\centering
\caption{Distribution of papers across different fields. NP indicates the percentage of papers in each field. The `Top 10-20\%' represents the percentage of the most cited papers within each area.
Biochemistry includes Genetics and Molecular Biology;  Pharmacology includes Toxicology and Pharmaceutics; Immunology includes Microbiology; and Biological Sciences includes Agricultural Sciences. The fields associated to each paper were obtained from the OpenAlex dataset.}
\label{tab:paper_distribution}
\begin{tabular}{|l|c|c|c|}
\hline
\textbf{Field} & \textbf{ NP (\%)} & \textbf{Top 20\%} & \textbf{Top 10\%} \\ \hline
Engineering & 40.4\% & 40.6\% & 38.9\% \\ \hline
Materials Science & 34.5\% & 33.6\% & 34.6\% \\ \hline
Energy & 8.0\% & 12.1\% & 13.7\% \\ \hline
Biochemistry & 5.0\% & 3.3\% & 2.7\% \\ \hline
Chemistry & 4.9\% & 4.8\% & 4.6\% \\ \hline
Physics and Astronomy & 1.9\% & 0.7\% & 0.8\% \\ \hline
Medicine & 1.8\% & 1.4\% & 1.4\% \\ \hline
Environmental Science & 1.4\% & 2.1\% & 2.1\% \\ \hline
Chemical Engineering & 0.7\% & 0.5\% & 0.6\% \\ \hline
Immunology & 0.4\% & 0.3\% & 0.2\% \\ \hline
Neuroscience & 0.4\% & 0.1\% & 0.1\% \\ \hline
Pharmacology & 0.2\% & 0.2\% & 0.1\% \\ \hline
Biological Sciences & 0.2\% & 0.1\% & 0.1\% \\ \hline
\end{tabular}
\end{table}

The dataset includes several types of information, \textit{viz.} the Digital Object Identifier (DOI) of each paper, the publication year, title, abstract, and the complete text of the paper. With DOI numbers we used the OpenAlex API \citep{priem2022openalex} to retrieve a time series of the citation count accumulated by each paper in the years following its publication. Papers lacking either the abstract or the main text were excluded to ensure the completeness of the data. Furthermore, for the latter part of our study, which involves prediction based on the full text, we segmented the full texts into their respective sections as demarcated by XML tags in the raw data. To preserve the original structure and semantic richness of the texts, we refrained from additional preprocessing steps, such as the removal of special characters or stop words. This approach was intended to maintain textual integrity and ensure the consistency of the inputs across the dataset.

\subsection{Target Variable Creation}

    The objective of this step was to construct the target variables (also known as labels) that will be predicted by the machine learning models. Our study concentrated on two target metrics indicative of a paper's citation performance: ACC (\emph{Accumulative Citation Count}) and YCC (\emph{Yearly Citation Count}). For ACC, we computed the cumulative citation count for each paper at several intervals post-publication. Papers were then ranked by their citation totals, with the top 20\% threshold used to define the boundary for high citation performance. Accordingly, for a specific paper and year pairing, ACC is assigned a value 1 if the paper falls within the top 20\% citation bracket for its publication year cohort, and 0 otherwise. YCC was defined similarly, with the calculation focusing on the citation count for each paper within a specified year after publication. The top 20\% are included in the high-performance set, with a YCC value of 1 assigned to those within this citation range, and 0 to the remainder. We also categorized the dataset into four groups: papers within the top 20\% for ACC versus the bottom 80\%, and papers within the top 20\% for YCC versus the bottom 80\%. To enable equitable comparisons across papers published in different years, we included a ``\emph{years ahead}'' control variable, facilitating an adjusted evaluation of citation performance as a function of time elapsed since publication. This control aids in juxtaposing, for instance, the citation counts of a paper two years after publication in 2014 with another two years post-publication in 2016.

Some limitations in our method are worth highlighting. First, citation frequency may be affected by temporal fluctuations in topic relevance and external factors unrelated to the intrinsic quality of the paper~\citep{amancio2012three}. Additionally, the method does not account for potential biases such as author reputation or institutional prestige, which could influence citation counts independently of a paper's merit. Moreover, papers that hover near the threshold of the top 20\% demarcation are susceptible to category shifts due to minor fluctuations in their citation numbers. 

\subsection{Feature Extraction Through Text Embeddings}

Textual content is converted into a machine-readable format, specifically into dense numerical vectors using various embedding models. These models are designed to capture semantic and syntactic features and encode them as vectors. The following embedding techniques were used:

\begin{itemize}
\item \emph{Generative Pre-trained Transformer (GPT)}: this model leverages multiple language tasks, such as translation, question-answering, and text completion, to learn language nuances. Using a transformer architecture, which employs an attention mechanism to identify relationships and context among words in both directions, these models can understand subtle linguistic details. We employed the GPT-3.5 ADA text embedding (text-embedding-ada-002)~\citep{DBLP:journals/corr/abs-2005-14165}, accessible through OpenAI's API, to embed the abstracts and full text of papers, yielding vectors with 1536 dimensions.

\item \emph{InferSent}: Developed by Facebook AI Research, InferSent \citep{DBLP:journals/corr/ConneauKSBB17} uses a Bi-LSTM network to process text in both directions, integrating a max pooling layer to capture the most salient features. This yields a comprehensive embedding of each sentence with a length of 4096.

\item \emph{Sentence-BERT}: Building on BERT's \citep{devlin2018bert} bidirectional capabilities, Sentence-BERT (SBERT) \citep{reimers2019sentence} employs a siamese network structure. Each segment of the network processes a sentence through BERT and follows with a pooling layer that aggregates features, leading to closely situated embeddings for similar sentences and divergent embeddings for differing ones. The resulting embeddings have 768 dimensions.

\item \emph{Term Frequency-Inverse Document Frequency (TFIDF)}: the approach embeds texts based on term occurrence frequencies~\citep{havrlant2017simple}. It balances term frequency in a document against its inverse frequency across the entire corpus, thus decreasing the weight of high-frequency words. We applied Singular Value Decomposition (SVD) \citep{stewart1993early} to reduce the high dimensionality of the TFIDF vectors, producing compressed embeddings that retain essential information. The final vector dimensions using SVD was set to 4096, matching InferSent with the highest size.

\item \emph{Universal Sentence Encoder (USE)}: the method employs transformer and deep averaging network (DAN) architectures to encode sentences~\citep{cer2018universal}. The transformer model is computationally intensive, but it excels at capturing the nuance of sentence meanings. The DAN model, on the other hand, averages word and bi-gram embeddings for efficiency. We utilized the DAN variant of USE, which outputs embeddings with a dimensionality of 512.
\end{itemize}

\subsection{Machine Learning Classification Algorithms}

\textcolor{black}{In the prediction phase, we employed four machine-learning algorithms to investigate how different models and embedding techniques perform. Since we did not focus on performance, no extensive hyperparameter optimization was made~\citep{feurer2019hyperparameter,yang2020hyperparameter}. The dataset was shuffled, with 30\% reserved for testing and 70\% used for training. We decided not to use cross validation given that it is time-consuming for more complex machine learning models in the context of the used dataset. The following models were used:}

\begin{itemize}
	\item \emph{Random Forest}: The Random Forest algorithm~\citep{biau2016random} operates on the principle of ``bagging'', which aggregates the outcomes of numerous decision trees to enhance predictive accuracy and control overfitting. Each tree is constructed using a different bootstrap sample from the dataset and a random subset of features, introducing diversity. A forest with more trees is typically more robust due to this de-correlation between individual trees. We used an ensemble of 1000 trees with default hyperparameters for classification tasks, with the final prediction determined by majority voting.

	\item \emph{XGBoost}: XGBoost~\citep{vital2022comparative,chen2016xgboost}, or eXtreme Gradient Boosting, is a sophisticated iteration of Gradient Boosted Machines (GBMs). It builds trees sequentially, each compensating for the weaknesses of its predecessors. The use of regularization techniques such as L1 and L2 and tree pruning helps prevent overfitting. Notable for its handling of missing data, computational efficiency, and performance, we configured XGBoost with 1000 trees for our experiments.

    \item \emph{K-Nearest Neighbors (KNN)}: this is an instance-based learning method that performs classification or regression by analyzing the 'k' most similar instances (neighbors) in the training data~\citep{guo2003knn}. It uses a distance metric to determine proximity, with the choice of 'k' affecting the model's bias-variance tradeoff. A smaller 'k' can lead to overfitting, while a larger 'k' may reduce sensitivity to the local structure of the data. In our study, we set k to 5.

    \item \emph{Logistic Regression}: this method is a straightforward yet effective model for binary classification~\citep{vital2022comparative,menard2002applied}. It applies a logistic function to estimate probabilities, which are then mapped to class labels based on a threshold, typically 0.5. Due to its simplicity and efficiency, it serves as a valuable baseline for evaluating the performance of more complex models.
    
\end{itemize}

\subsection{Performance Evaluation}

To quantify the classification quality, we selected two performance metrics: Area Under the Receiver Operating Characteristic Curve (AUC-ROC) and Area Under the Precision-Recall Curve (AUC-PR). AUC-ROC is normally used for binary classification tasks, including for link prediction in the context of paper citation~\citep{vital2022comparative}. Given our context of an unbalanced dataset, where the positive class is rare, we also use AUC-PR which is sensitive to the performance of the positive class. The metrics are defined as follows: 

\begin{itemize}
	\item \emph{Area Under the Receiver Operating Characteristic Curve (AUC-ROC)}: AUC-ROC is a measure of model performance in binary classification that plots the true positive rate (TPR) against the false positive rate (FPR) at various threshold settings. An AUC-ROC value of 1 indicates perfect discrimination, while 0.5 suggests no discriminative power — equivalent to random guessing.

    \item \emph{Area Under the Precision-Recall Curve (AUC-PR)}: AUC-PR is particularly informative for imbalanced datasets. It plots precision (the ratio of true positives to all positive predictions) against recall (the ratio of true positives to all actual positives). Unlike metrics dependent on a fixed threshold, such as the F1 score, AUC-PR integrates over all possible thresholds, providing insight into model performance across a range of precision and recall values. A higher AUC-PR indicates a model's better capability to identify positive cases effectively, regardless of dataset balance.
    
\end{itemize}

\subsection{Datasets}
We employed two distinct datasets to evaluate our models:

\begin{itemize}
    \item \emph{Balanced dataset}: To avoid class bias, we created a balanced dataset by undersampling \citep{liu2008exploratory} the negative class, ensuring an equal number of instances from the top 20\% of papers (by citation count) and the remaining 80\%. This approach allows us to assess the models in a controlled environment, providing insights into their performance in an ideally balanced scenario.

	\item \emph{Skewed dataset}: Reflecting the natural distribution in academic citations, this dataset contains a disproportionate number of papers not within the top 20\% citation bracket. This skewed distribution presents a more challenging and realistic test for the models, probing their ability to cope with class imbalances that are typical in real-world data.
\end{itemize}

\section{Results and Discussion}

In Section \ref{subsectionresultsA}, we analyzed the effectiveness of diverse textual embedding models and machine learning classification algorithms using paper abstracts as input. This involved a thorough comparative examination of outcomes for the target variables, YCC and ACC. In Section \ref{subsectionresultsB} we analyzed the embedding classification, specifically investigating the classification performance over the years post-publication. Lastly, in Section \ref{subsectionresultsC}, we used the most effective combination of classification algorithm and embedding model to encode the full text of papers.

\label{chapter:Results and Discussion}

\subsection{Embedding and Classification Models Analysis}
\label{subsectionresultsA}

\begin{figure}[H]
\centering
\includegraphics[scale=.45]{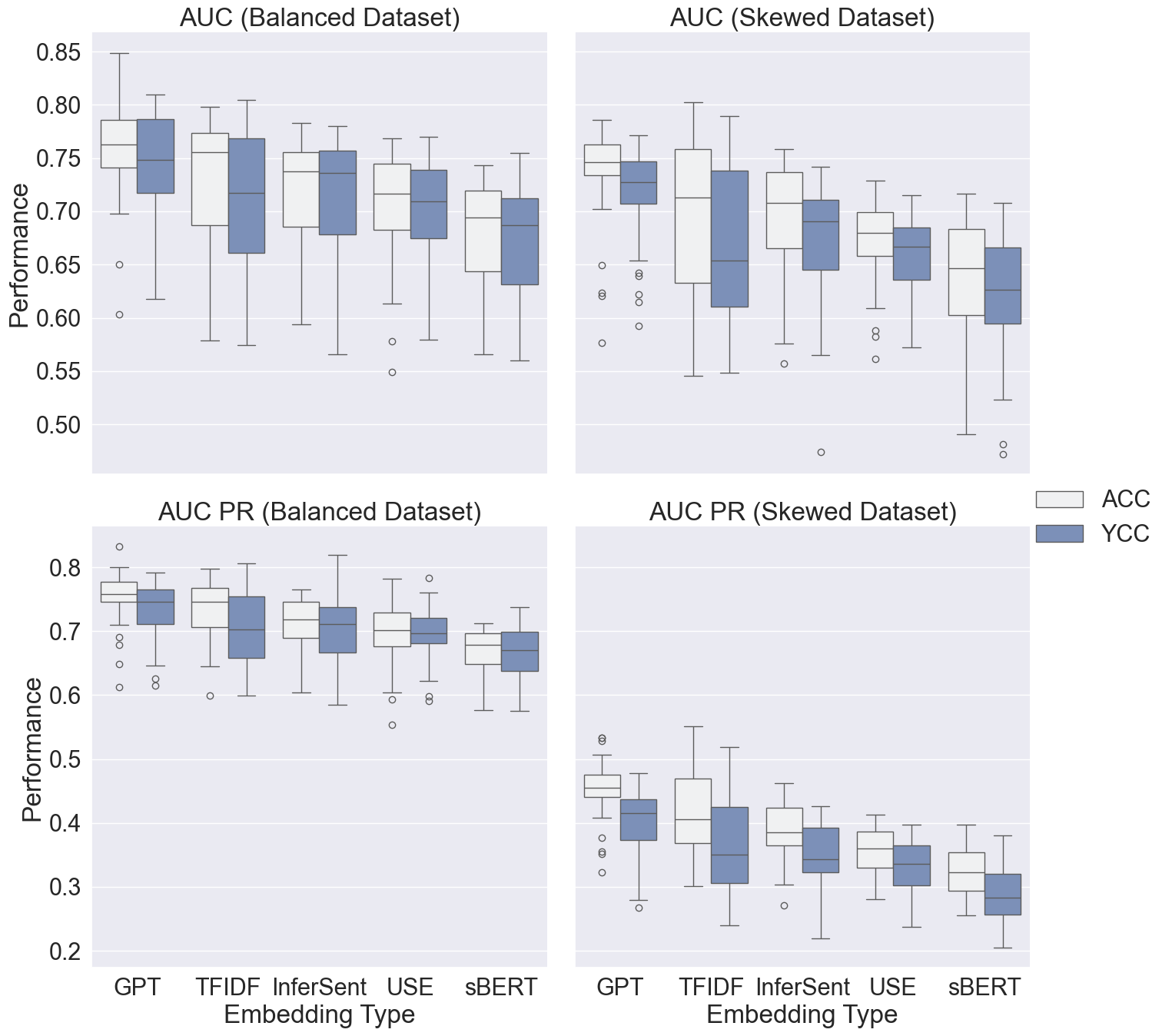}
\caption{Comparative performance of \emph{text embedding models} on citation prediction. Box plots illustrating the effectiveness of different embedding techniques in predicting the top 20\% cited academic papers over the years using different classification algorithms. Performance is evaluated using (AUC-ROC) and (AUC-PR) across the balanced and skewed datasets. ACC denotes the prediction of papers in the top 20\% of accumulative citations over the years, while YCC refers to the prediction of papers with the top annual citations in a specific year. The results highlight the variance in model performance, with implications for model selection in citation analysis tasks. \label{figure:embedding}}
\end{figure}

Figure \ref{figure:embedding} compares the performance with AUC-ROC and AUC-PR for the metrics ACC and YCC using various embedding models and the balanced and skewed datasets. Results were derived by aggregating, using the median, the performance of various classification algorithms across multiple prediction years. ACC consistently surpasses the YCC metric, which varies to a greater extent as indicated by its broader interquartile range. This observation suggests potential challenges in the model's ability to predict a paper's impact based solely on annual performance metrics. In contrast, more robust and consistent results were obtained for the cumulative performance metrics across different setups, indicating a higher likelihood of predicting a paper's impact when considering its cumulative citations over time. Models trained on balanced datasets exhibited superior performance, which should be expected because balanced training allows the model to grasp the subtleties of the classes more efficiently. In contrast, skewed datasets introduce biases, leading to an overemphasis on features from the predominant class and inadequate training for underrepresented classes. AUC-ROC for balanced datasets ranges between 0.69 and 0.76, compared to 0.63 and 0.75 for skewed datasets. For AUC-PR, the values vary from 0.67 to 0.76 for balanced and from 0.28 to 0.46 for skewed datasets. In both metrics, GPT consistently exhibits the highest median, while sBERT has the lowest. It is worth noting that the decrease in AUC-ROC for skewed datasets is less significant compared to the reduction observed in AUC-PR. This is understandable, as the former metric is less sensitive to imbalances than the latter, which is more affected by the precision-recall tradeoff in skewed distributions. Ultimately, since real-world datasets are typically skewed, they inherently present greater predictive challenges, leading to increased variability and diminished performance compared to an ideal, balanced dataset.

Among the embedding models, GPT consistently outperformed the others in all scenarios. It achieved a median AUC-ROC of 0.76 for balanced datasets and 0.75 for skewed, with its highest performance being 0.85 AUC for balanced datasets. Remarkably, the second-best performing model was TFIDF, a simpler approach compared to others. Its AUC median fluctuated between 0.72 to 0.76 for balanced and 0.65 to 0.71 for skewed datasets, peaking at an AUC of 0.8. The small difference (0.05) between GPT and TFIDF is notable, given that TFIDF essentially functions as an advanced term counter; yet it secured the second position, surpassing models that utilize transformer networks and attention mechanisms. This indicates that not only semantic and sentence meanings are crucial for predictions, but also the specific terms used in a paper carry significant predictive weight. While this performance with TFIDF may seem surprising, one has to consider that certain terms inherently draw more attention, while others may deter readers and subsequent citations, regardless of the paper's overall quality. This opens up opportunities for further research into the specific terms that contribute to a paper's success over time, including the potential identification of terms that maintain their relevance regardless of temporal factors.

\begin{figure}[H]
\centering
\includegraphics[scale=.45]{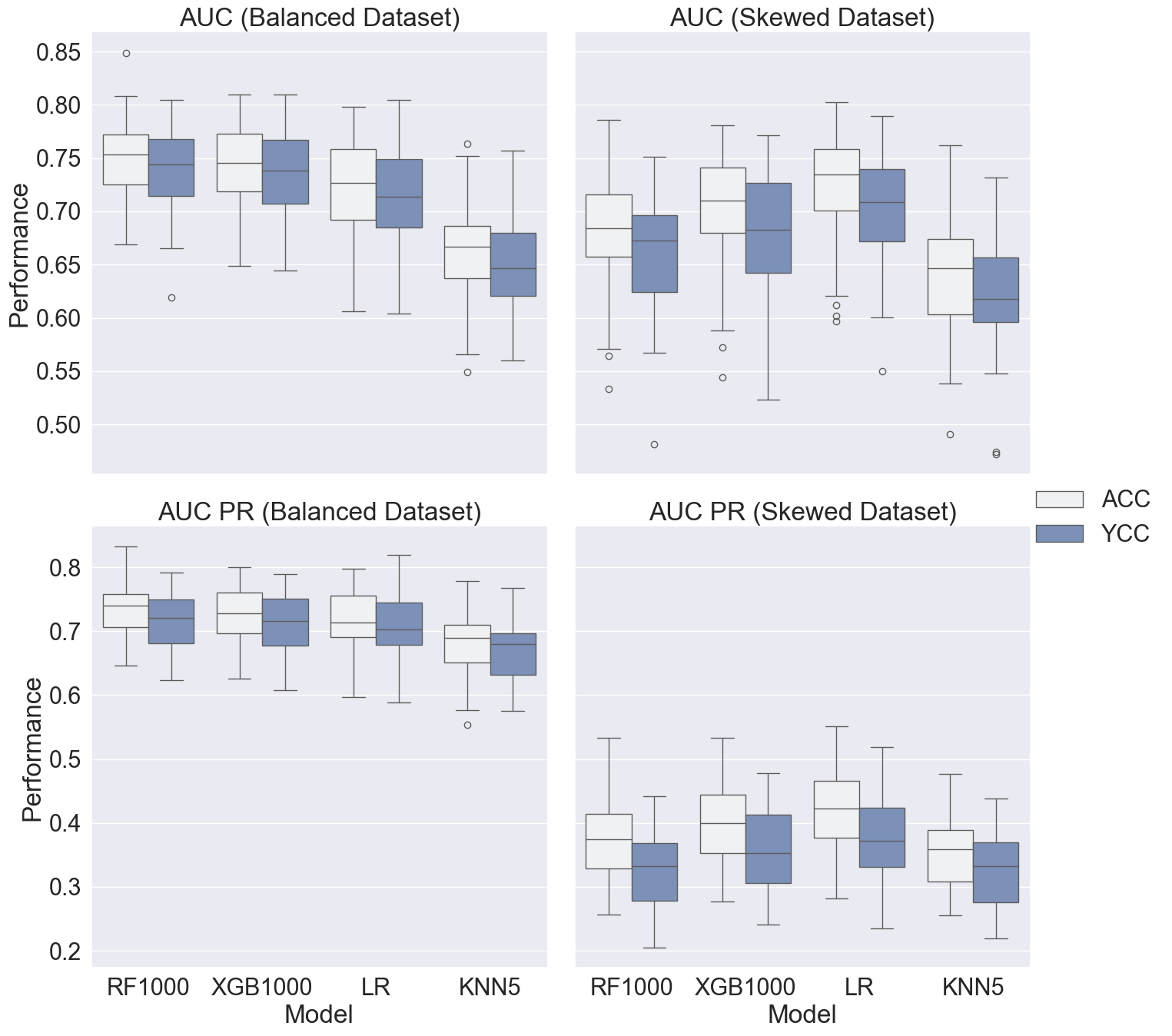}
\caption{
\textcolor{black}{Comparative performance of \emph{machine learning classification algorithms} on citation prediction. The box plots illustrate the effectiveness of the different embedding techniques in predicting the top 20\% cited academic papers over the years using different classification algorithms. Performance is evaluated using the Area Under the ROC Curve and of the Precision-Recall Curve (AUC-PR) for the balanced and skewed datasets. ACC denotes the prediction of papers in the top 20\% of accumulative citations over the years, while YCC refers to the prediction of papers with the top annual citations in a specific year. The results highlight the variance in model performance, with implications for model selection in citation analysis tasks. }\label{figure:model}}
\end{figure}

We now analyze the performance achieved by traditional classifiers. The box plots in Figure \ref{figure:model} refer to the performance indicators AUC-ROC and AUC-PR for ACC and YCC using the classification models RF1000 (Random Forest with 1000 trees), XGB1000 (XGBoost with 1000 estimators), LR (Logistic Regression), and KNN5 (K-Nearest Neighbors with 5 neighbors). As in the previous analysis, higher performances are observed for ACC and for the balanced dataset. One exception to the trend occurred with logistic regression (LR) for which AUC  was practically the same for the two datasets. Apart from KNN, all models seem to perform in the same range for the balanced dataset, with AUC medians between 0.71 and 0.74, and AUC-PR between 0.71 and 0.73. For the skewed dataset, AUC medians ranged from 0.67 to 0.73, while AUC PR varied from 0.33 to 0.42. Note that with the skewed dataset not only the performances were lower but there was a higher uncertainty in the results. For the balanced dataset, random forest had the highest AUC median of 0.75 when used to predict ACC and 0.74 for YCC, and for AUC PR it had 0.74 and 0.72, respectively.

\subsection{Performance of the embedding models over the years post-publication}
\label{subsectionresultsB}

\begin{figure}[H]
\centering
\includegraphics[scale=.4]{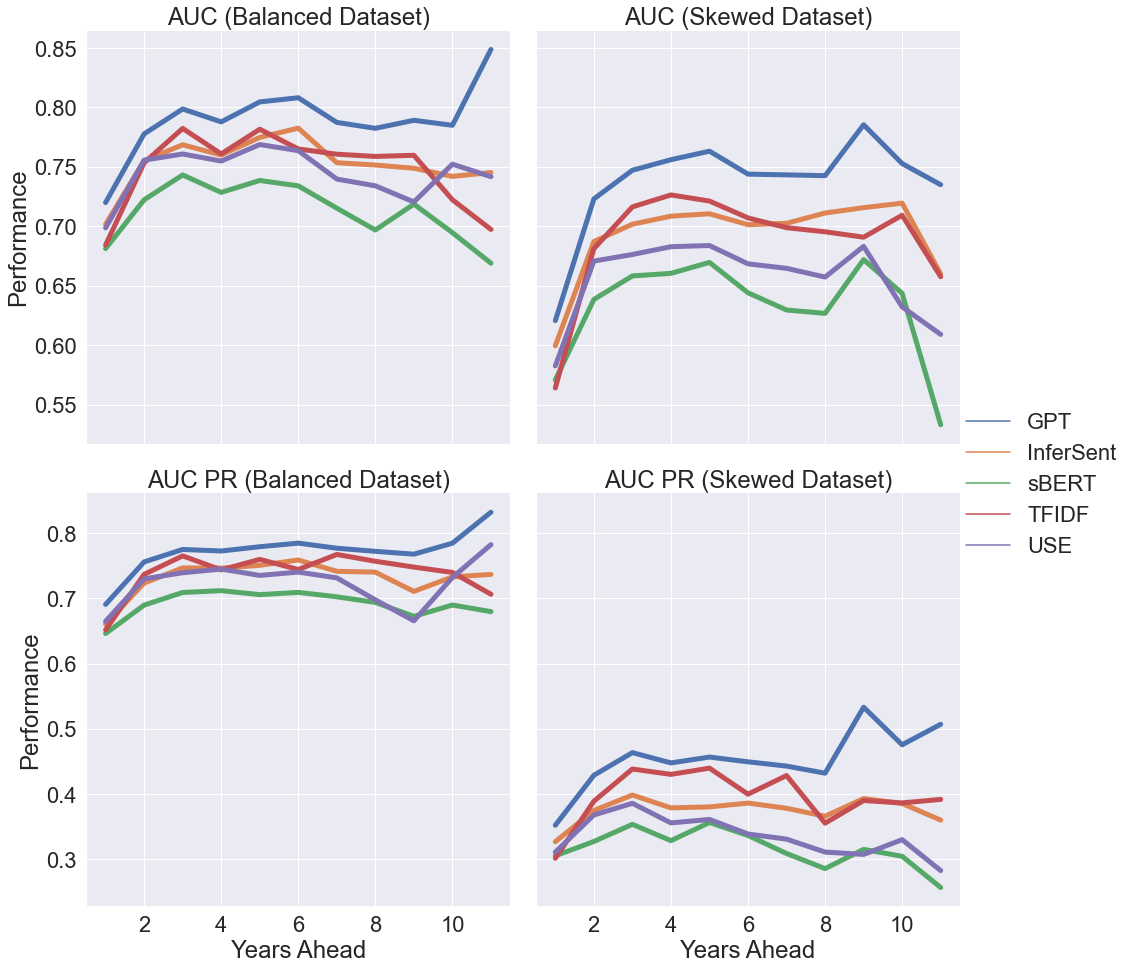}
\caption{Yearly performance trends of text embedding models for predicting top 20\% accumulative citations in academic papers over an 11-Year horizon using {AUC-ROC} and AUC-PR metrics on balanced and skewed datasets. Here we used the ACC metric as a measure of performance.\label{figure:best_acc}}
\end{figure}

Figure \ref{figure:best_acc} shows AUC-ROC and AUC-PR for both datasets using a random forest model with 1000 trees (RF1000) and various text embedding models 
to predict ACC with the accumulative number of citations over a time frame of 11 years. {
The rationale for employing only the random forest model stems from its consistent outperformance relative to others when applied to a balanced dataset}. 

\begin{figure}[H]
\centering
\includegraphics[scale=.4]{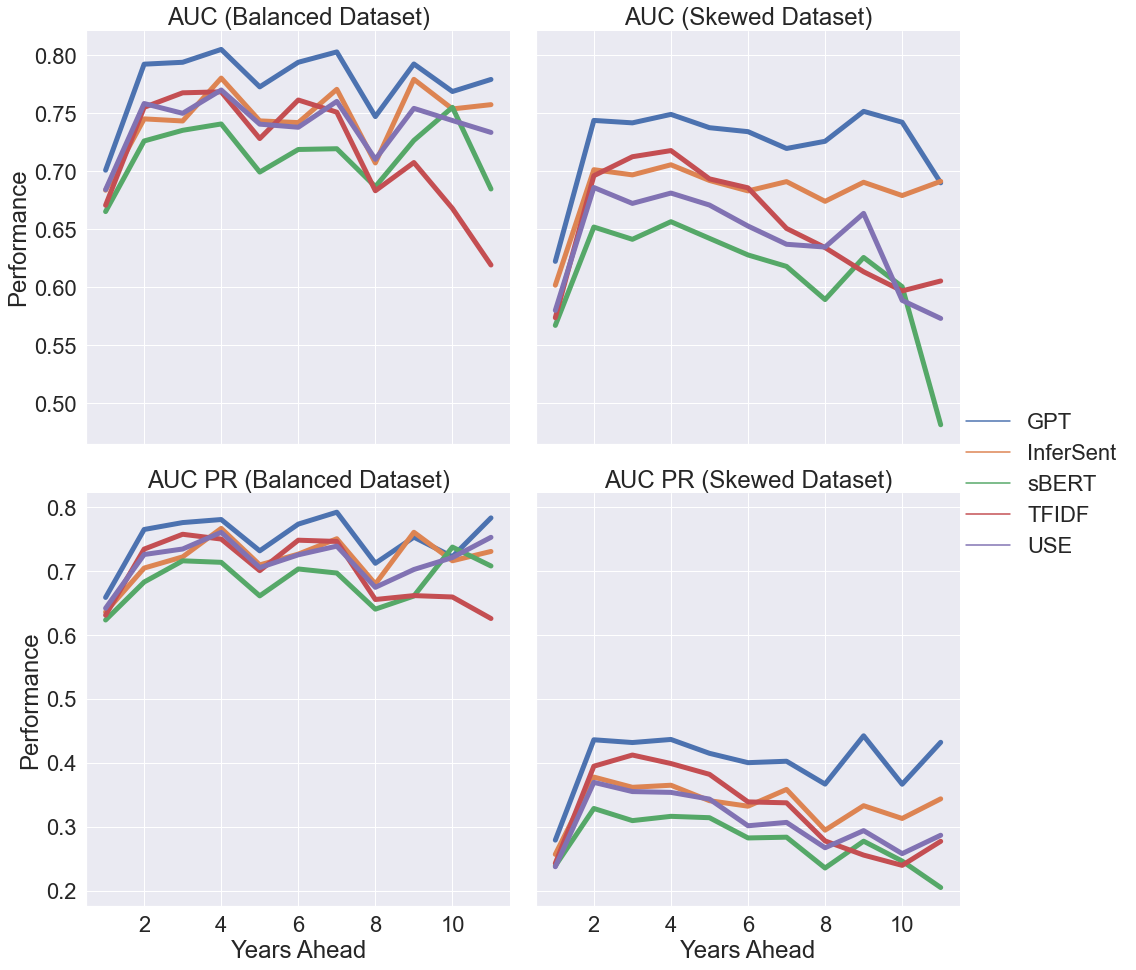}
\caption{Yearly performance trends of text embedding models for predicting top 20\% annual citations in academic papers over an 11-Year Horizon using AUC-ROC and AUC-PR metrics on balanced and skewed datasets. Here we used the YCC metric as a measure of performance.\ \label{figure:best_ano}}
\end{figure}

The predictive performance is lower in the first year following the paper publication for two reasons: the number of citations is normally small and it depends on whether the paper was published early or at the end of the year. After the third year, the performance seems to stabilize for most embedding models. The GPT model consistently outperforms the others, reaching its highest AUC-ROC score in the eleventh year.  After the sixth year, we observe a downturn in the performance of the TFIDF, USE, and sBERT models, which may be due to the challenges of making long-term predictions. One should note that the difference in performance between GPT and TFIDF when considering only the random forest model is higher than in Figure 1 where the median values of several machine learning models were used. For AUC-PR, the performance remains steady over time, suggesting a sustained balance between precision and recall. GPT is the top performer, while sBERT lags behind, with TFIDF and InferSent in the second and third places. For the skewed dataset, the variability in performance is more pronounced, and unlike the balanced dataset, there is no clear pattern over the years. The AUC scores generally improve until the fourth or fifth year and then slightly decrease, followed by an unpredictable rise and fall. With AUC PR, the GPT model, which leads in performance, shows a general upward trend. This could be attributed to its advanced architecture and the contemporary data used in its training, which may give it an edge over the other models. For the YCC metric, Figure \ref{figure:best_ano} shows a similar behavior to that for ACC with some more fluctuation. Again, GPT is the highest performer, and performance is normally higher for the balanced dataset. Overall, the results show that the balanced dataset yields the highest performance, with performance increasing after 2-3 first years.

\textcolor{black}{In summary, our findings support existing theories highlighting the importance of content in determining a paper's visibility and citation potential~\citep{wang2013quantifying}. While these results may not be applicable across all scientific fields, we found content to be a good predictor of citations, independent of traditional scholarly metrics like author reputation. Not all text features, however, are useful for predicting impact; for example, text complexity has been shown to be uncorrelated with a paper's influence~\citep{lu2019analyzing}. Additionally, studies have shown that the impact of papers within the same journal can be comparable, even if they belong to distinct subtopics~\citep{chacon2020comparing}. This underscores the idea that impact is not a simple function of any single textual feature like topic, complexity, or sentiment~\citep{yousif2019survey}.
}

\subsection{Prediction using full text or abstract only}
\label{subsectionresultsC}

\begin{figure}[H]
\centering
\includegraphics[scale=.4]{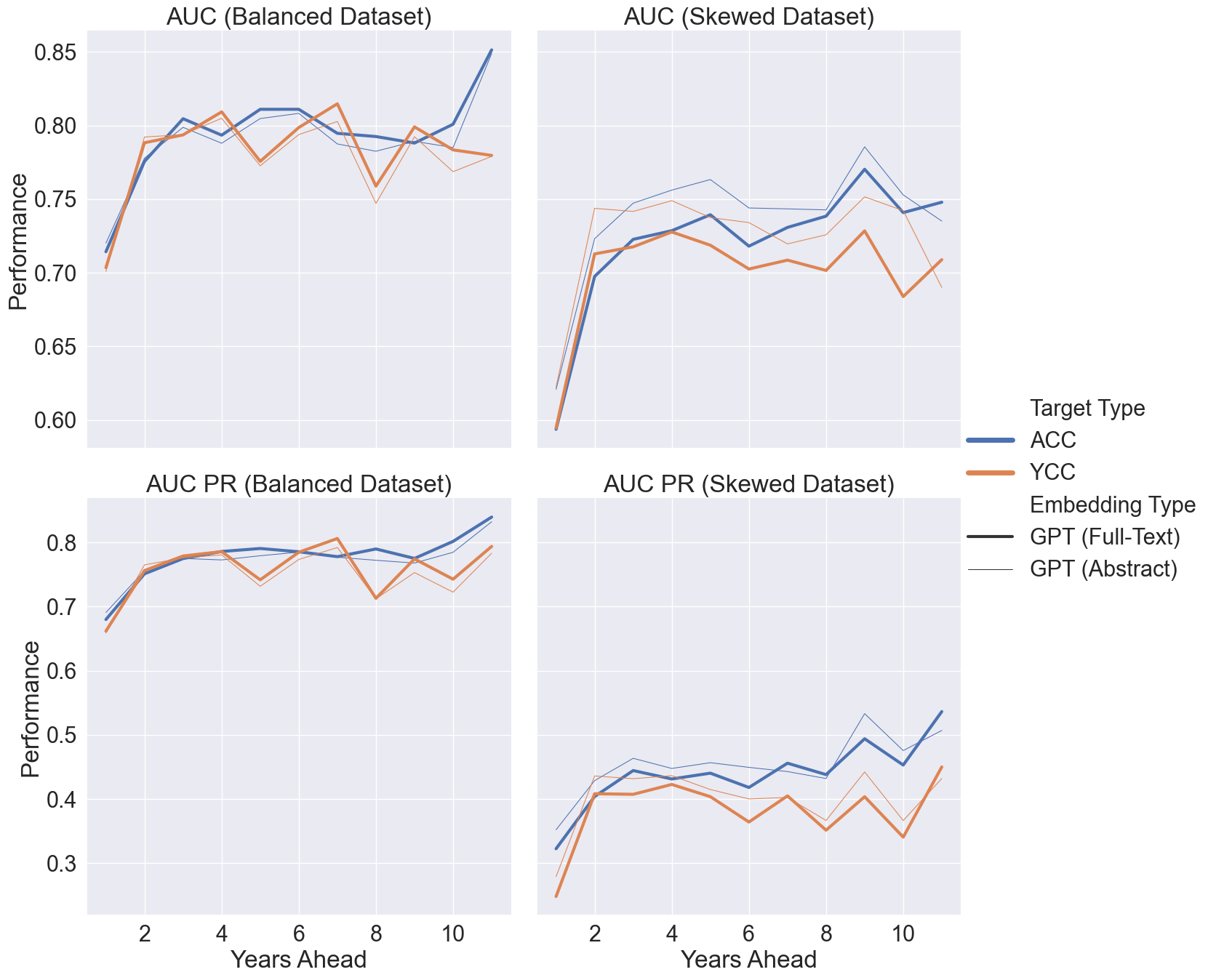} 
\caption{Comparative analysis of random forest with 1000 trees and GPT embeddings for Full-Text versus abstract-only for academic papers performance over 11 years post-publication using AUC and AUC PR performance metrics across balanced and skewed datasets. \label{figure:gpt_full_paper}}
\end{figure}

In the preceding sections, we used the abstracts of the papers as key pieces of information for predicting their impact. To verify whether processing the full text would lead to a higher performance in predicting impact, we employed the best-performing classification algorithm, Random Forest with 1000 trees, and the most effective text embedding model, GPT. The papers were segmented into sections using XML tags, embedded each section separately with GPT, and then averaged these embeddings for each paper. This process was used to predict our two key metrics, ACC and YCC, across various years following publication. The results obtained with full text are depicted in Figure \ref{figure:gpt_full_paper}. The bold line represents the full-text analysis and the thinner line represents the abstract-only approach. Interestingly, incorporating the entire text into our predictive model resulted in only a modest increase in performance for the balanced dataset for both the YCC and ACC metrics. In contrast, when examining the skewed dataset, the simpler abstract-only embeddings actually delivered better results most of the time, with the AUC metric showing the most significant difference between the two approaches. This suggests that while full-text analysis may offer slight advantages in an idealized dataset, for more realistically distributed (skewed) datasets, abstracts alone provide enough information to yield superior predictive outcomes. This outcome also hints at the possibility that including full text may introduce extraneous data that could hamper the predictive accuracy. A deeper investigation might involve analyzing predictions using specific sections of the papers to determine if focusing on particular parts of the text could improve performance.

\section{Conclusion}
\label{chapter:conclusion}

We explored the use of machine learning and text embedding to predict the impact of scientific papers published in the journal ACS Applied Materials and Interfaces based on their citation counts. In this methodology, the only relevant factors for impact were the quality of the work and the topics covered in a paper. The most efficient combination was found using a Random Forest classifier and the GPT embedding model, achieving an average accuracy of approximately 80\%  for the metric ACC (accumulated citation count) in predicting papers among the top 20\% most cited. This level of accuracy is noteworthy, especially considering that factors such as the reputation of authors, institutions, countries, and citation patterns were not taken into account. In other words, the methodology captured the topics and syntax-semantic aspects of the text, which typically correlate with text quality. Three other findings are worth noting: i) the prediction accuracy remained similar whether only the abstracts or the full texts were analyzed; ii) a much simpler, less computationally expensive embedding based on TFIDF—essentially capturing paper topics—performed closely to GPT. 
iii) the results remain robust when considering a larger or smaller fraction of top-cited papers in the balanced dataset (see Appendix). 
\textcolor{black}{
Considering the results observed in the ACS AMI journal dataset, these findings suggest that a robust predictive power of impact can be achieved by identifying the paper's topics and `reading' the abstract.} This conclusion aligns with the dependence of the impact factor on the topics~ 
and with expectations from editors who emphasize the relevance of the abstract in screening papers for a journal. 

\textcolor{black}{Predicting impact can be useful for several reasons, including its role in allocating resources to more promising research. Additionally, prediction systems can help identify disruptive research that might otherwise be overlooked. However, there are ethical implications to consider when using such systems in practice. One potential issue is the bias inherent in the data used to train these systems. For example, if a particular journal consistently associates higher impact with certain methodologies, both the prediction system and human reviewers might undervalue alternative research approaches. Bias can also arise if the system takes into account factors like the author or the university, potentially disregarding lesser-known researchers. Furthermore, another concern with prediction systems is their lack of interpretability, which can make it challenging to understand the reasons behind a paper's projected impact.}

In future works, it would be beneficial to investigate how certain sections of academic papers contribute to predictions. One promising approach could be to integrate author impact metrics, like citation counts or h-index, into the predictive models.
Other methodologies could be explored, such as integrating large language models (LLMs) with graph-based analyses~\citep{onan2023hierarchical,stella2020forma,pan2024unifying,correa2019word,amancio2012using}.
Most importantly, it would also be interesting to consider not only accuracy but also the bias and interpretability of the system when making predictions.

\section*{Acknowledgments}

This work was supported by CNPq and FAPESP (2018/22214-6) (Brazil). The authors also thank the American Chemical Society for providing us with the full text of the ACS AMI journal. Filipi N. Silva thanks AFOSR \#FA9550-19-1-0391 for the financial support.

\bibliographystyle{apalike}

\bibliographystyle{abbrvnat}

\newpage

\section*{Appendix: Robustness Analysis}

\textcolor{black}{To evaluate the robustness of the results, we assessed how the performance vary when considering different numbers of top papers. A test was conducted using papers published between 2012 and 2020, with citations tracked until 2023. We used an 80/20 split for the training and testing datasets, proportionally based on the publication year. The classification model was a neural network with hidden layers containing 32 and 16 neurons, a dropout layer with a 50\% probability, and a sigmoid function in the output layer for predictions. We employed a 5-fold cross-validation method to statistically ensure the model’s performance across different shares of top papers, including the top 10\%, 20\%, 30\%, and 40\%, to test the model's robustness. In this analysis, we used the GPT model.}

\begin{figure}
\centering
\includegraphics[scale=.4]{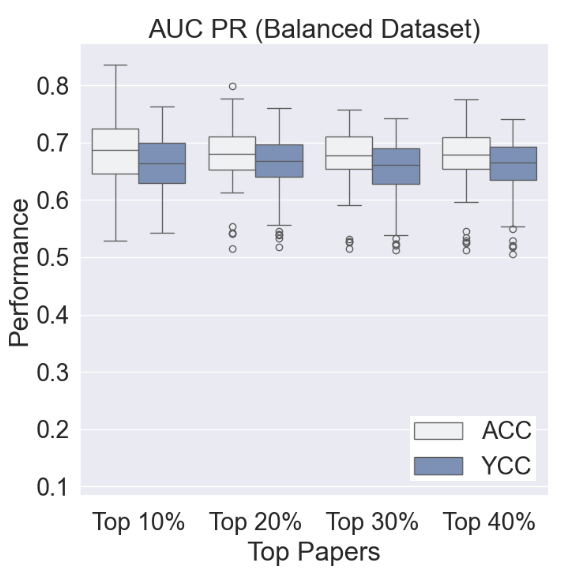}
\caption{Performance on citation prediction. The box plots illustrate the effectiveness in predicting the top 10, 20, 30 and 40\% cited papers over the years using different classification algorithms.  ACC denotes the prediction of papers in the top accumulative citations over the years, while YCC refers to the prediction of papers with the top annual citations in a specific year. \label{figure1}}
\end{figure}

\begin{figure}
\centering
\includegraphics[scale=.27]{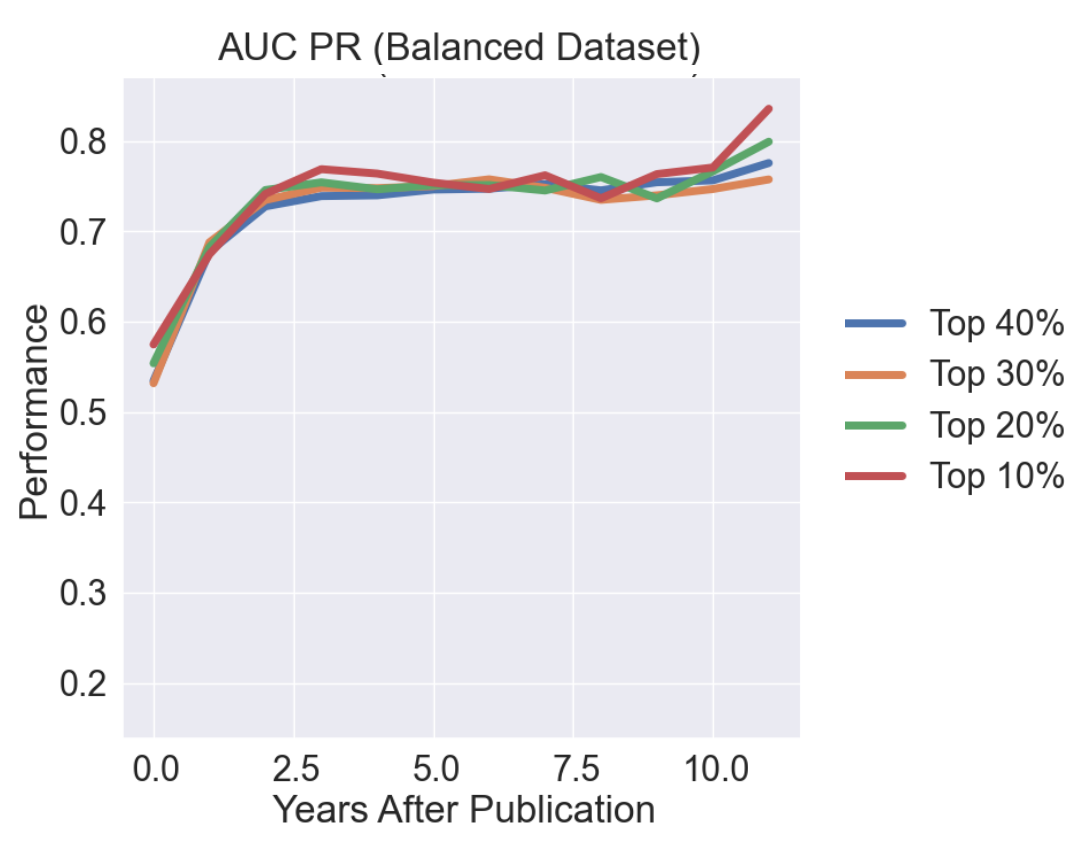}
\caption{Yearly performance trends of GPT model for predicting top 10-40\% annual citations in academic papers over an 11-Year Horizon using  AUC-PR metric on the balanced dataset. Here we used the YCC metric as a measure of performance. \label{figure2}}
\end{figure}

\textcolor{black}{When considering the results based on aggregation across multiple prediction years (as discussed in Section \ref{subsectionresultsA}), we found that in the balanced dataset, the performance exhibits low variation when different numbers of papers are considered as top papers. This result is illustrated in Figure \ref{figure1}. This behavior is particularly evident in the balanced dataset, reinforcing the importance of balancing classes.
A similar behavior was observed when the analysis was conducted for specific years following the publication of the paper (as discussed in Section \ref{subsectionresultsB}). The performance remained robust when considering the balanced dataset, as shown in Figure \ref{figure2}.}

%

\newpage

\end{document}